%% file: kubat1.tex
\documentclass[natbib]{svmult}

\usepackage{mathptmx}       
\usepackage{helvet}         
\usepackage{courier}        
\usepackage{type1cm}        
\usepackage{makeidx}         
\usepackage{graphicx}        
\usepackage{multicol}        
\usepackage[bottom]{footmisc}

\makeindex             

\usepackage{amsmath,amssymb}
\usepackage{url}

\newcommand{\pul}{\ensuremath{\frac{1}{2}}}
\newcommand{\japul}{\ensuremath{\frac{3}{2}}}

\newcommand{\pderiv}[2]{\frac{\partial #1}{\partial #2}}
\newcommand{\pderivl}[2]{{\partial #1}/{\partial #2}}
\newcommand{\nelec}{\ensuremath{n_\text{e}}}
\newcommand{\melec}{\ensuremath{m_\text{e}}}

\newcommand{\zav}[1]{\left(#1\right)}

\newcommand{\Jiri}{Ji\v{r}\'{\i}}
\newcommand{\Kubat}{Kub\'at}

\newcommand{\ATLAS}{{\tt ATLAS9}}
\newcommand{\SYNTHE}{{\tt SYNTHE}}
\newcommand{\TLUSTY}{{\tt TLUSTY}}
\newcommand{\SYNSPEC}{{\tt SYNSPEC}}

\begin{document}

\title*{Basics of the NLTE physics}
\titlerunning{Basics of the NLTE physics}
\author{{\Jiri} {\Kubat}}
\institute{{\Jiri} {\Kubat} \at Astronomick\'y \'ustav AV \v{C}R,
Fri\v{c}ova 298, 251 65 Ond\v{r}ejov, Czech Republic,
\email{kubat@sunstel.asu.cas.cz}}
%
%
\maketitle

\abstract{
Basic assumptions of the NLTE approximation in stellar atmospheres are
summarized.
The assumptions of thermodynamic equilibrium, local thermodynamic
equilibrium (LTE), and non-LTE (NLTE) are compared.
It is emphasized that LTE is a poor approximation if radiative
transitions dominate in stellar atmospheres.
The equations of kinetic equilibrium and methods of their solution are
discussed.
}

\section{Introduction}

Let us assume that using some spectrograph and telescope we have
observed a stellar spectrum, and that we have reduced it using some data
reduction software.
Such spectrum is ready for its analysis.
If we decide to do a model atmosphere analysis of this spectrum we need
to proceed in several steps using different kind of software.
First, we have to calculate a model atmosphere (spatial distribution of
temperature, density, and occupation numbers), which can be done using
the assumption of local thermodynamic equilibrium (LTE), for example
using the popular code {\ATLAS} \citep{Kurucz:1993},
or without this restrictive assumption allowing the atomic level
population to differ from their equilibrium values (so called NLTE),
which can be done using the code {\TLUSTY} \citep{Hubeny:1988}.
With the model atmosphere calculated we then have to determine
detailed emergent spectrum.
This can be done using spectral synthesis codes (like {\SYNTHE}
associated with {\ATLAS} or {\SYNSPEC} associated with \TLUSTY), which
calculate the formal solution of the radiative transfer equation for as
many frequencies as we need, so an accurate theoretical emergent stellar
radiation is obtained.
The final step of the analysis is comparison of the theoretical spectrum
with observations, which can also be done using a computer code.

It is clear that for each step of this procedure different computer
codes are used.
Unfortunately, some users focus mostly to the last step (comparison of
observations with the theoretical spectrum) and
they are satisfied when any theoretical spectrum fits observations.
They do not care about details of model atmosphere calculation and
associated assumptions.
However, a good fit of a simplified model does not necessarily mean that
the simplification is acceptable.
Taking into account more accurate physics generally leads to better
models.
Switching from the assumption of LTE to the assumption of NLTE is such
step towards better models.
Particular steps in the above mentioned spectrum analysis are influenced
differently if LTE or NLTE are assumed.
The initial step of model atmosphere calculation strongly depends on the
assumption LTE/NLTE.
Spectral synthesis is dependent on the initial step, since data input to
the code is more complex for the NLTE case.
The comparison step is the same if we used the NLTE model or not.

In this paper we shall discuss the physical and microscopic bases of
the assumption of NLTE in a greater detail.

\section{Microscopic distributions}

Solution of the stellar atmosphere problem can be considered as the
determination of spatial dependence of basic macroscopic quantities
\citep[see e.g.][and references
therein]{Mihalas:1978,Hubeny:Mihalas:2014}.
It can also be understood \citep[following][]{Hubeny:1976} as searching
for basic microscopic distributions,
namely, the momentum distribution (which is equivalent to distribution
of velocities of all particles),
distribution of particle internal degrees of freedom (i.e. populations
of atomic excitation stages),
and distribution of internal degrees of freedom of the electromagnetic
field (which is the radiation field for all frequencies and directions).
Several different approximations may be applied to determination of
these distributions.

\subsection{Thermodynamic equilibrium}

In thermodynamic equilibrium, all these three distributions have their
equilibrium values.
A system in thermodynamic equilibrium experiences no changes when it is
isolated.
By relaxation time we understand time necessary for the system to return
back to equilibrium from a perturbed state.
Macroscopic changes have to be slower with respect to relaxation time.

In thermodynamic equilibrium, the radiation field has the Planck
distribution and velocities of particles follow the Maxwell
distribution.
Individual atomic energy levels (states) are populated according to
Boltzmann distribution
\begin{equation}\label{boltzmann}
\frac{n_{i,j}^\ast}{n_{0,j}^\ast} = \frac{g_{i,j}}{g_{0,j}}
e^{-\frac{\chi_{0i,j}}{kT}}.
\end{equation}
When the equation \eqref{boltzmann} is applied to ground levels of two
successive ions, it describes the ionization equilibrium (the Saha
equation),
\begin{equation}\label{saha}
\frac{n^\ast_{0,j+1}}{n^\ast_{0,j}} = \frac{2g_{0,j+1}}{g_{0,j}}
\frac{1}{\nelec} \zav{\frac{2\pi \melec kT}{h^2}}^\japul
e^{-\frac{\chi_{I,j}}{kT}}.
\end{equation}
In these equations,
$T$ is temperature,
$\melec$ is the electron mass,
$\nelec$ is the electron number density,
$n_{i,j}^\ast$ is the equilibrium population and
$g_{i,j}$ is the statistical weight of the level $i$ of the ion $j$
(similarly for $i=0$, the ground level),
$\chi_{0i,j}$ is the excitation energy of the level $i$ from the ground
level,
$\chi_{I,j}$ is the ionization energy of the ground state of the ion
$j$,
$h$ is the Planck constant,
and $k$ is the Boltzmann constant.
Expressions for these distributions as well as discussion about
influence of individual processes on maintenance of equilibrium
distributions are presented also in \cite{nice2}, together with
references to literature.

However, thermodynamic equilibrium can not be used for description of
the stellar atmosphere.
As it is evident from observations of stellar spectra, the stellar
radiation does not obey the Planck distribution.

\subsection{Local thermodynamic equilibrium (LTE)}

To describe the stellar atmosphere, which is the transition from the
dense and  hot star to almost void and cold insterstellar medium, we
have to introduce the possibility of depth dependent density and
temperature in the stellar atmosphere.
This also allows the radiation to escape from the stellar atmosphere.

To preserve the advantages of the thermodynamic equilibrium, we
introduce the {\em local thermodynamic equilibrium}, where we assume
Maxwell distribution of particle velocities and Saha-Boltzmann
distribution of excitation and ionization states to be valid locally
using local values of temperature $T$ and electron density $\nelec$.
Thus the atomic energy level populations are calculated after equations
\eqref{boltzmann} and \eqref{saha}.
The radiation field is no more in equiulibrium (does not have Planck
distribution).
It has to be determined by solution of the radiative transfer equation,
however, the assumption of local thermodynamic equilibrium allows to use
the equilibrium source function, which is equal to the Planck function.

\subsection{Kinetic equilibrium (NLTE or non-LTE)}

The assumption of kinetic equilibrium \citep[see][]{Hubeny:Mihalas:2014}
is more general than the assumption of local thermodynamic equilibrium.
In kinetic equilibrium we determine the radiation field by solution of
the radiative transfer equation and we use the exact source function.
Populations of individual atomic levels are calculated using equations
of kinetic equilibrium (which include the non-local influence of
radiation on level populations), and the particle velocities are assumed
to be equilibrium (Maxwell distribution).
This approximation is usually called non-LTE or NLTE, which looks like
negation of equilibrium.
This is evidently not true, since in this ``non-equilibrium'' there are
still particle velocities in equilibrium.
Even more confusing is the expansion of the ancronym LTE, which gives
``non-local thermodynamic equilibrium'', which is also far from the
reality.
More exactly, by non-LTE we understand \emph{any} state that departs
from LTE. 
Therefore, it is better to follow \cite {Hubeny:Mihalas:2014} and call
this approximation {\em kinetic equilibrium}, in contrast to the
commonly used (but inexact) term statistical equilibrium.
Since the ancronym NLTE is commonly used for this approximation, we
shall use it also in this paper.

How close the distribution of excitation states comes to the equilibrium
one, depends on a balance between radiative and collisional processes.
If the collisional processes dominate, then the distribution is close to
the equilibrium one.
On the other hand, if radiation processes dominate, then the
distribution may differ from the equilibrium one significantly.
In NLTE, no assumption is made about the source function, it is
consistently calculated from actual opacities and emissivities.

Generally, collisional processes tend to establish equilibrium (since
particles have equilibrium velocity distributions), radiative process
tend to destroy it if radiation field is not in equilibrium (which is
the case in stellar atmospheres).

A detailed discussion of individual processes may be found in
\citet[][see Table 1]{nice2}.

\subsection{Basic comparison of LTE and NLTE}

Both in the local thermodynamic equilibrium (LTE) and in the kinetic
equilibrium (NLTE) the maxwellian (equilibrium) distribution of particle
velocities is assumed.
Frequent elastic collisions between particles preserve equilibrium
velocity distribution.
Inelastic collisions tend to destroy it, however, if the number of
elastic collisions between two inelastic collisions is very large, any
deviation from equilibrium caused by inelastic collision is quickly
compensated and equilibrium is preserved.
In other words, for equilibrium velocity distribution it is necessary
that the relaxation time is much shorter that time between inelastic
collisions.
Since this condition is fulfilled in most cases in stellar atmospheres,
we assume in the following that the particle distribution is maxwellian
(equilibrium).

The basic difference between LTE and NLTE is the behavior of calculated
atomic level populations.
The approximation of LTE allows their relatively very simple and fast
calculation using the Saha-Boltzmann distribution (Eqs.~\ref{boltzmann}
and \ref{saha}).
The NLTE approximation takes into account the fact that the level
populations are influenced by the radiation field.
In this case, the populations have to be determined by the equations of
kinetic equilibrium (Eqs.~\ref{obecese} or \ref{statese} later in this
paper).

It is the balance between several types of microscopic processes, which
tells us how exact (or inexact) the assumption of LTE is.
The validity of the LTE approximation is determined by a competition
between collisional and radiative processes.
The radiative processes tend to destroy the equilibrium distributions
(if the radiation field is not in equilibrium), while the inelastic
collisions tend to establish equilibrium distribution of excitation
states, however, provided that the elastic collisions are even more
frequent than the inelastic ones.
As collisional processes we may consider processes, where more than one
particle (electron, ion, but not a photon) take part.
A special case is radiative recombination, which is a radiative process
(radiation is emitted), but a collision of two particles (namely an ion
and an electron) must happen.
For a more detailed list of microscopic processes see Table~1 in
\cite{nice2}.

The necessary microscopic condition for the validity of LTE is the
condition of {\em detailed balance}, which means that rates of all
processes are balanced by rates of their reverse processes.
If the equilibrium particle velocities distribution holds, then also
collisional excitation and ionization processes are in detailed balance.
However, the radiative processes are in detailed balance only if the
radiation field is in equilibrium, i.e. if it obeys the Planck
distribution.
If the mean intensity of the radiation field $J(\nu) \ne B(\nu)$, we can
not reach detailed balance in radiative transitions.
Since in stellar atmospheres the radiation field {\em is generally not}
in equilibrium, detailed balance can not be achieved if radiative
transitions dominate, and, consequently, LTE is a poor approximation
there.
Illustrative discusssion may be found in \cite{Mihalas:Athay:1973}.

\section{Equations of kinetic equilibrium}

The general form of kinetic equilibrium equations for the state (energy
level) $i$ is \citep[][Eq. 5-48, see also \citealt{Hubeny:Mihalas:2014}]
{Mihalas:1978}
\begin{equation}\label{obecese}
\pderiv{n_{i}}{t} + \nabla \cdot \zav{n_{i} \vec{v}} =
\sum_{\substack{l=1 \\ l\ne i}}^L \zav{n_{l} P_{li} - n_{i} P_{il}}
\end{equation}
($i=1,\dots,L$),
where $L$ is the total number of energy levels considered, and $n_i
P_{il}$ is the number of transitions per time unit from the state $i$ to
the state $l$ (which includes both line and continuum transitions).
The quantity $P_{il}$ is the transition rate\footnote{Note that
sometimes $n_i P_{il}$ is called the rate.
Here we prefer to use the notation consistent with the textbook of
\cite{Hubeny:Mihalas:2014}.}.
It can be expressed as a sum, $P_{il} = R_{il} + C_{il}$, where $R_{il}$
is the rate of the transition caused by absorption or emission of
radiation, and $C_{il}$ is the rate of the transition caused by a
collision with a neighbouring particle, mostly with an electron.
While the collisional rates depend only on local values of electron
density $\nelec$ and temperature $T$, the radiative rates depend also on
the radiation field, which is non-local.
Detailed expressions for both collisional and radiative rates are listed
in another part of this book \citep{maria1}.
Multiplying the equation \eqref{obecese} by mass of the corresponding
particle and summing all such equations over all energy levels of all
species we obtain the continuity equation
$\pderivl{\rho}{t} + \nabla \cdot \left( \rho \vec{v} \right)=0$.

For the case of a static atmosphere ($\vec{v}=0$) in a steady state
($\pderivl{}{t}\rightarrow 0$), we may neglect both the time derivative
and the advective term in the equation \eqref{obecese} and we obtain the
time independent set of the kinetic equilibrium equations for static
medium,
\begin{equation}\label{statese}
\sum_{\substack{l=1 \\ l\ne i}}^L \left(n_{l} P_{li} - n_{i} P_{il}
\right) = 0.
\end{equation}
However, this equation is commonly used not only for static atmospheres,
but the advective term is very often neglected also for expanding
stellar envelopes.

\subsection{Equilibrium level populations}

If we want to measure how much the population numbers differ from their
LTE values, we use the departure coefficients ($b$-factors)
$b_{i,j}=n_{i,j}/n_{i,j}^\ast$ \citep{Menzel:1937}, which relate actual
($n_{i,j}$) and equilibrium ($n_{i,j}^\ast$) populations of the level
$i$ of the ion~$j$.

There are two ways how to define the equilibrium populations.
The first one is the natural one.
The equilibrium populations in the whole stellar atmosphere are
calculated using Saha-Boltzmann distributions using temperature and
density distributions obtained also for the assumption of local
thermodynamic equilibrium.

On the other hand, \cite{Mihalas:1978} uses the quantity $n_{i,j}^\ast$
in a different meaning.
The LTE populations are calculated with respect to the population of the
ground level of the next higher ion using the equation (5-14) in
\citet{Mihalas:1978}, which follows from Eqs.~\eqref{boltzmann} and
\eqref{saha},
\begin{equation}
n^\ast_{i,j} = n_{0,j+1} \nelec \frac{g_{ij}}{g_{0,j+1}}
{\pul \zav{\frac{h^2}{2\pi mkT}}^\japul}
e^{-\frac{\chi_{Ij}-\chi_{ij}}{kT}}.
\end{equation}
Here $n_{0,j+1}$ is the {\em actual} population of the ground level of
the next higher ion, which does not need to be an LTE value.
This definition of the LTE value of population numbers reflects the fact
that the radiative recombination is an equilibrium process, which causes
that the ionization transitions from the highest levels of each ion $j$
are close to detailed balance.

\subsection{System of kinetic equilibrium equations}

For each atom, the kinetic equilibrium equations \eqref{statese} can be
written as
\begin{equation}
n_i \sum_{\substack{l=1 \\ l\ne i}}^L \zav{R_{il}+C_{il}} -
\sum_{\substack{l=1 \\ l\ne i}}^L n_l \zav{R_{li}+C_{li}} = 0.
\end{equation}
where $i=1,\dots,L$ and $L$ is the total numbers of atomic energy levels
considered.
For each atom, the system of equations is linearly dependent.
We have to replace one of the equations by a `closing equation', which
sets the total atom particle density and thus makes the system linearly
independent.
For model atmosphere calculations we can use the charge conservation
equation, which compares the total charge of all ions included in
kinetic equilibrium with the charge of electrons.
Alternatively, the particle conservation equation may be used.
This equation uses the total number density and thus fixes the total
number of particles involved.
If more atoms are included in the equations of kinetic equilibrium,
additional abundance equation is used as a closing equation.
Kinetic equilibrium equations for each independent atom have to be
closed by one such equation.
These equations are listed in \cite{nice2}.

The full set of kinetic equilibrium equations can be formally written as
\begin{equation}\label{esesys}
{\cal A} \cdot \vec{n} = {\cal B},
\end{equation}
where $\vec{n} = (n_1, \dots, n_L)$ is a vector of all populations,
the matrix ${\cal A}$ contains all rates included (it is called
the rate matrix) and the right hand side vector ${\cal B}$ is 0 except
rows with conservation laws.

\subsection{Solution of the system of kinetic equilibrium equations}

For given ${\cal A}$ and ${\cal B}$, \eqref{esesys} is a set of linear
equations with a straightforward solution.
However, the rate matrix ${\cal A}$ depends on the radiation field.
As a consequence, we have to solve the set of equations of kinetic
equilibrium {\em simultaneously} with the radiative transfer equation
for all frequencies condsidered.
This is a nonlinear system of equations.
There is a simple iteration scheme, which consists of subsequent
solution of the radiative transfer equation for given opacity and
emissivity (the formal solution of the radiative transfer equation) and
solution of the equations of kinetic equilibrium for given radiation
field.
However, this iteration scheme, known as the $\Lambda$-iteration, does
not work for stellar atmospheres.
More precisely, it converges so slowly that it is unusable for media
which are optically thick at some frequency.
Illustrative examples can be found in \citet{Auer:1984}.
Consequently, different iteration schemes must be used.
It may be either the multidimensional Newton-Raphson method
\citep[introduced to modeling of stellar atmospheres by][and usually
referred to as the {\em complete linearization
method}]{Auer:Mihalas:1969}
or the {\em accelerated $\Lambda$-iteration technique} \citep[introduced to
radiative transfer by][]{Cannon:1973:1,Cannon:1973:2}, which
differs from the ordinary lambda iteration by using the approximate
lambda operator.
This operator allows simultaneous solution of the equations of kinetic
equilibrium together with the approximate radiative transfer equation
with a subsequent exact solution of the radiative transfer as a
correction term.
Useful reviews of accelerated lambda iteration methods were published by
\citet{Hubeny:1992,Hubeny:2003}.
These solution methods may be used also for the generalized problem of
calculation of full NLTE model atmosphere, where additional equations of
radiative and hydrostatic equilibrium are simultaneously solved.

\section{Summary}

The ancronym NLTE (or non-LTE) denotes deviation from LTE which in
stellar atmospheres means that:
\begin{itemize}
\item the radiation field is not in thermodynamic equilibrium, its
intensity has to be determined from the radiative transfer equation,
\item excitation and ionization state of matter is not in thermodynamic
equilibrium, the level populations have to be determined from equations
of kinetic equilibrium,
\item equations of radiative transfer and kinetic equilibrium have
to be solved {\em simultaneously} because they are coupled,
\item particle velocities are in thermodynamic equilibrium, they obey
Maxwell distribution.
\end{itemize}
This approximation gives significantly better and more exact description
of reality than the ``standard'' approximation of LTE.

\begin{acknowledgement}
The author would like to thank Dr. Ewa Niemczura for inviting him to the
Spring School and he would also like to apologize her for the delay in
delivering manuscripts.
He is also grateful to both referees (Ivan Hubeny and the anonymous one)
for their invaluable comments.
This work was partly supported by the project 13-10589S of the Grant
Agency of the Czech Republic (GA \v{C}R).
\end{acknowledgement}

\input{casopisy}

\bibliographystyle{spbasic}
\bibliography{kubat,wrspec,proc}

\end{document}

%% file: casopisy.tex
\newcommand{\aj}{Astron. J.}
\newcommand{\aap}{Astron. Astrophys.}
\newcommand{\aapr}{Astron. Astrophys. Rev.}
\newcommand{\apj}{Astrophys. J.}
\newcommand{\apjs}{Astrophys. J. Suppl. Ser.}
\newcommand{\araa}{Ann. Rev. Astron. Astrophys.}
\newcommand{\jqsrt}{J. Quant. Spectrosc. Radiat. Transfer}
\newcommand{\mnras}{Mon. Not. Roy. Astron. Soc.}